\documentclass[12pt]{iopart}
\usepackage{graphicx}
\usepackage[english]{babel}

\begin{document}

\title{Dynamics of perturbations in Gurzadyan-Xue cosmological models}
\author{G. V. Vereshchagin\footnote{%
e-mail address: \texttt{veresh@icra.it}} and G. Yegorian\footnote{%
e-mail address: \texttt{gegham@icra.it}}}
\address{ICRANet, P.le della Repubblica 10, I65100 Pescara, Italy and ICRA, Dip.
Fisica, Univ. \textquotedblleft La Sapienza\textquotedblright , P.le A. Moro
5, I00185 Rome, Italy}
\date
\maketitle

\begin{abstract}
Perturbation theory within Newtonian approximation is presented for cosmological models with varying physical constants. Analytical solutions for perturbations dynamics are obtained for each Gurzadyan-Xue model with pressureless matter and radiation. We found that perturbations grow during entire expansion within GX models, including curvature- and vacuum-dominated stage when they cease to grow in the standard cosmological model.
\end{abstract}

\section{Introduction}

The formula for dark energy, derived by Gurzadyan and Xue predicts the
observed value for the density parameter of the dark energy without any free
parameters \cite{GX}. The formula reads:

\begin{equation}
\rho _{GX}=\frac{\pi }{8}\frac{\hbar c}{L_{p}^{2}}\frac{1}{a^{2}}=\frac{\pi 
}{8}\frac{c^{4}}{G}\frac{1}{a^{2}},  \label{rhoLambda}
\end{equation}%
where $\hbar $ is the Planck constant, the Planck length is $L_{p}=\left(
\hbar G\right) ^{\frac{1}{2}}c^{-3/2}$ and $c$ is the speed of light, $G$ is
the gravitational constant. Here $a$ is the upper cutoff in computation of
vacuum fluctuations and we take it to be the scale factor of the Universe,
although other choices are possible \cite{DG}. According to Zeldovich \cite%
{Zel67}, the vacuum energy (\ref{rhoLambda}) corresponds to the cosmological
term

\begin{equation}
\Lambda _{GX}=\frac{8\pi G\rho _{GX}}{c^{2}}=\pi ^{2}\left( \frac{c}{a}%
\right) ^{2},  \label{Lambda}
\end{equation}

\bigskip

In General Relativity all the quantities in (\ref{rhoLambda}) and (\ref%
{Lambda}) are constants, except for the scale factor, which is a function of
cosmic time. Therefore, adopting Gurzadyan-Xue (GX)\ scaling (\ref{rhoLambda}%
) one has to consider possible variation of physical constants\footnote{%
Generally speaking, the vacuum energy is known to be constant only in
Minkowski spacetime. However, assuming it is constant one has unique
connection between the vacuum energy density and the cosmological term and
can interpret one of them as a fundamental constant of physics.}.

Simple cosmological models, based on above ideas we proposed in \cite{Ver06}%
. In particular, models with varying cosmological term, speed of light and
gravitational constant were considered. Qualitative analysis of solutions in
these models \cite{Ver06a} revealed some interesting features, in particular
the presence of separatrix in the phase space of solutions. This separatrix
divides the space of solutions into two classes:\ Friedmannian-like with
initial singularity and non-Friedmannian solutions which begin with nonzero
scale factor and vanishing matter density. Each solution is characterized by
a single quantity, the density parameter which is defined in the same way as
in the standard cosmological model:

\begin{equation*}
\Omega _{m}=\frac{\mu _{0}}{\mu _{c}},
\end{equation*}

where $\mu $ is matter density, $H$ is Hubble parameter, $\mu _{c}=\frac{%
3H_{0}^{2}}{8\pi G_{0}}$ is critical matter density, and index "$0$" refers
to the values today. Separatrix in all models is given by the density
parameter $\Omega _{m}\approx 2/3$ that indicated some hidden symmetries
between the models, although cosmological equations look very different.
Later in \cite{khach} the origin of these symmetries was discovered and
attributed to existence of invariants in GX models.

Note, that the energy-momentum conservation does not lead to mass
conservation of the usual matter in models with varying constants. This can
be interpreted as creation of matter due to variation of constants. As a
consequence, the usual power law $\mu \propto a^{-3}$ does not hold in
expanding Universe. In general $\mu(a)$ is obtained by solving
cosmological equations.

Analytical solutions for all GX\ models both for matter density and the
scale factor are obtained in \cite{Ver06b}. It turns out that the most
simple solutions for the scale factor are again those of separatrix. In one
model it is exponential, in the others they are polynomials.

Simple GX models were generalized to include radiation in \cite{Ver06b},
where analytical solutions were also obtained. The main difference in that
case is absence of separatrix for matter density solutions in models
with radiation:\ all solutions
(with exception of model with varying cosmological constant) begin with
vanishing matter density in contrast to Friedmannian solutions. There is,
however, a particular density parameter in those models that divides
solution again into two classes: in the one class the matter density
increases in matter-dominated epoch, in the other it decreases.

Thus, GX models possess interesting and nontrivial features. The original
motivation for these cosmological models, however, is the fact that the
predicted dark energy density is close the value inferred from the set of current
observations. Along this line, it is of crucial importance to further
confront predictions of these models with observations. We have performed
likelihood analysis of supernovae and radio galaxies data within GX models
in \cite{Ver06c}. Such important characteristics of models such as age and
deceleration parameter were also computed. Our results show that models with
density parameter smaller than the separatrix value do not pass
observational constraints. This indicates the preference of high matter
densities: GX models with $2/3\leq \Omega
_{m}\leq 1$ pass these simple but important constraints. All these results
show that background dynamics of GX\ models is viable in front of recent
observations.

In this paper we turn to analysis of perturbations dynamics within GX
models. It is well known that the most stringent constraints on cosmological
parameters within $\Lambda $CDM models were obtained from observations of
large scale structure distribution of galaxies and clusters, as well as from
cosmic microwave background radiation anisotropies. We will consider simple
Newtonian perturbations, i.e. those with wavelengths well inside the
horizon. This allows, nevertheless, to understand the main differences with
respect to dynamics of perturbations in the Friedmannian models.

The paper is organized as follows. In sec. 2 GX cosmological models are
reviewed. In sec. 3 we present cosmological perturbation theory with varying constants in Newtonian approximation. In sec. 4 this theory in applied
to GX models. In sec. 5 discussion and conclusions follow.

\section{Gurzadyan-Xue cosmological models}

In this section we remind briefly the content of GX models, for detailed
discussion see \cite{Ver06b}.

Adopting the scaling (\ref{rhoLambda}) and postulating the validity of
Einstein equations when speed of light and gravitational constant vary with
time we obtain four simplest cases when the speed of light, the
gravitational constant or vacuum energy density vary with time in such a way
that (\ref{rhoLambda}) is satisfied. Below we provide cosmological equations
for each model. The content of the Universe is supposed to be a pressureless
fluid with mass density $\mu $ and energy density $\mu c^{2}$ as well as a
radiation field with energy density $\rho _{r}$ and pressure $\rho _{r}/3$.

We also assume that evolution of radiation energy density is described by
the following equation

\begin{equation*}
\rho _{r}=\rho _{r0}\left( \frac{a}{a_{0}}\right) ^{-4},
\end{equation*}%
which means there is no energy flows into radiation field. This assumption allows to avoid problems with anisotropies of cosmic microwave background and large scale structure data for models with varying constants \cite{Oph04},\cite{Oph05}.

All GX model with radiation are characterized by two density parameters: $\Omega _{m}$ and $%
\Omega _{r}=\rho_{r0}/(\mu_c c_0^2)$.

\subsection{Model 1, with radiation}

We suppose that neither the speed of light, nor the gravitational constant
change with time, but the vacuum energy density does, so $\rho _{GX}\propto
a^{-2}$. This is the unique case when the radiation does not couple to
matter \cite{Ver06b}.

This model is defined by the following cosmological equations \cite{Ver06b}

\begin{eqnarray}
H^{2}+\frac{k^{\prime }c^{2}}{a^{2}} &=&\frac{8\pi G}{3}\left( \mu +\frac{%
\rho _{r0}}{c^{2}}\left( \frac{a}{a_{0}}\right) ^{-4}\right) ,  \\
\dot{\mu}+3H\mu &=&\frac{\pi }{4G}\left( \frac{c}{a}\right) ^{2}H,
\end{eqnarray}%
where $\rho _{r0}$ is radiation energy density today, and effective
curvature is $k^{\prime }=k-\frac{\pi ^{2}}{3}$, $k=-1,0,1$. The first
equation looks identical to Friedmann equation without cosmological
constant. The second equation corresponds to continuity equation with the
source term due to vacuum energy variation. This term is positive during
expansion and may be interpreted as creation of matter.

\begin{figure}
	\centering
		\includegraphics{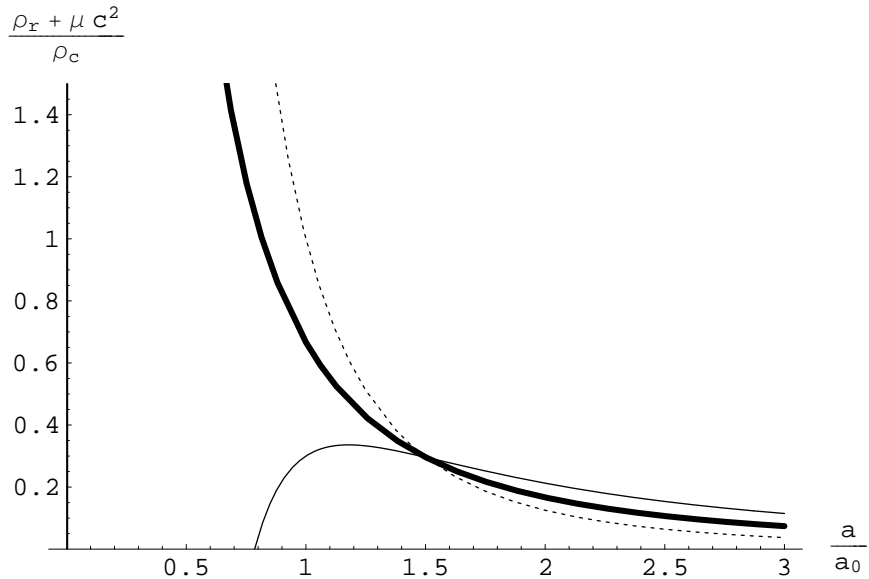}
	\label{fig1}
	\caption{Total energy density as a function of scale factor for model 1 with radiation. Here $a_{0}$ is scale factor today. Critical energy density is defined as $\protect\rho _{c}=%
\protect\mu _{c}c^{2}$. In this and subsequent figures black curve correspond to density parameter $\Omega _{m}=2/3$, dotted curve is for $\Omega _{m}=0.3$ while firm curve is for $\Omega _{m}=1$. Also for all figures $k=0$.}
\end{figure}

\begin{figure}
	\centering
		\includegraphics{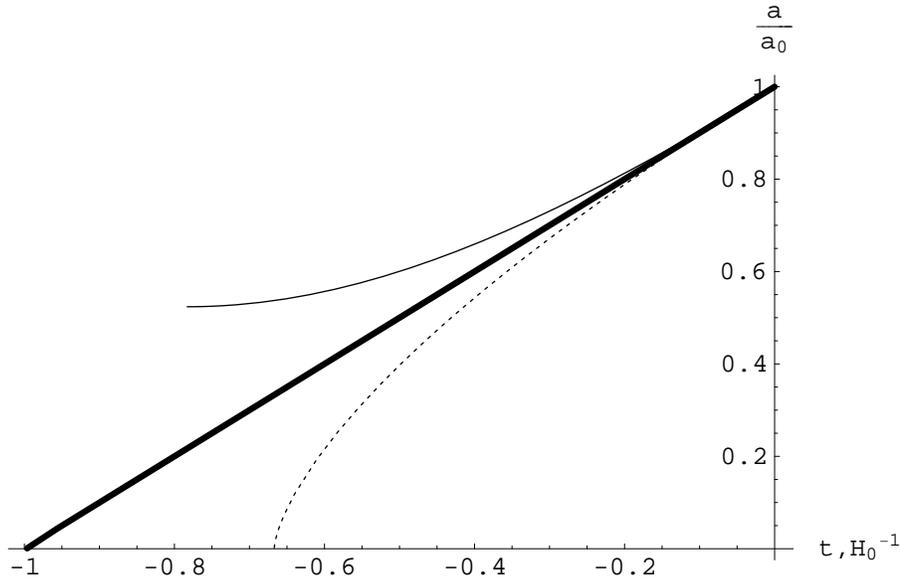}
	\caption{Scale factor as a function of time for model 1 with radiation. Here time is measured in inverse Hubble parameter today, for which we take $H_{0}=70$ km/(s Mpc), so $H_{0}^{-1}=14$
GYr for this and all subsequent figures.}
	\label{fig2}
\end{figure}

Solutions of these equations are represented at fig. \ref{fig1}, \ref{fig2}. Matter density
and energy density of pressureless fluid are related through constant speed
of light.


This model contains radiation-dominated (RD) stage which begins from
Friedmannian singularity if $\Omega _{m}<\Omega _{sep}$, where density
parameter for \emph{separatrix} is $\Omega _{sep}=\frac{2}{3}\frac{1-\Omega
_{r}}{1-k/\pi ^{2}}$, or with finite value of scale factor and vanishing
energy density if $\Omega _{m}>\Omega _{sep}$. Then RD stage is followed by matter-dominated (MD) stage. On last (kD) stage curvature dominates\footnote{More precisely, terms of the order $a^{-2}$ dominate in the first cosmological equation.}.

\subsection{Model 2, with radiation}

In this model the speed of light changes with time in such a way that vacuum
energy density (\ref{rhoLambda}) together with (\ref{Lambda}) remain constants, so
that the latter may be interpreted as cosmological term. The speed of light
increases during cosmic expansion:\ $c\propto a^{2}$. Our cosmological
equations are

\begin{eqnarray}
H^{2}-\frac{\Lambda ^{\prime }}{3} = \frac{8\pi G}{3}\left( \mu +\frac{\pi
^{2}\rho _{r0}}{\Lambda }\frac{a_{0}^{4}}{a^{6}}\right), \\
\dot{\mu}+H\left( \mu -\frac{\Lambda }{4\pi G}\right) = \frac{4\pi ^{2}\rho
_{r0}a_{0}^{4}}{\Lambda }\frac{H}{a^{6}}.
\end{eqnarray}%
where $\Lambda ^{\prime }=\Lambda \left( 1-3k/\pi ^{2}\right) $ is effective
cosmological constant.

\begin{figure}
	\centering
		\includegraphics{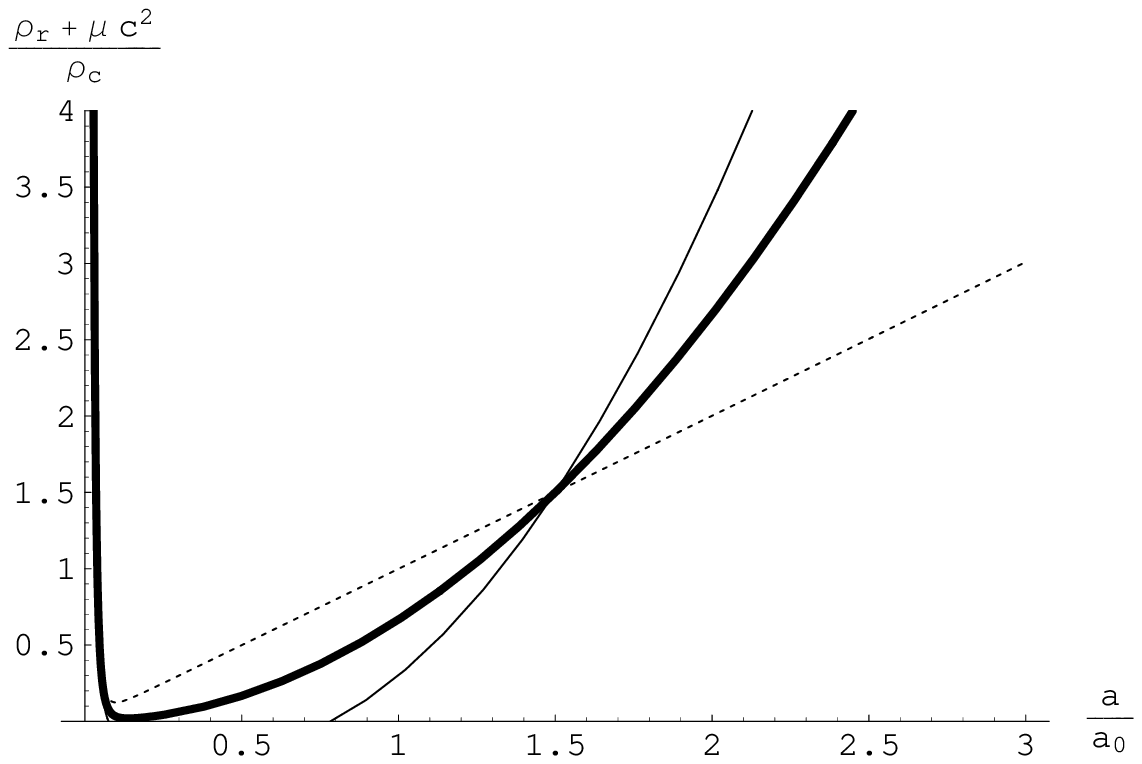}
	\caption{Total energy density as a function of scale factor for model 2 with radiation.}
	\label{fig3}
\end{figure}

\begin{figure}
	\centering
		\includegraphics{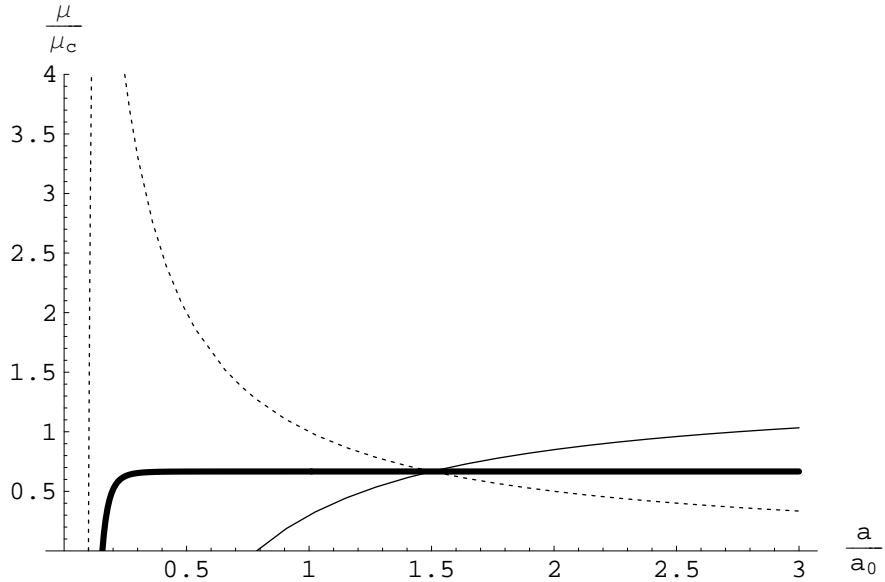}
	\caption{Matter density of pressureless component in units of critical density as a function of scale factor for model 2 with radiation. For solution with $\Omega_m>\Omega_{sep}$ the matter density has a maximum which is not shown here.}
	\label{fig4}
\end{figure}

\begin{figure}
	\centering
		\includegraphics{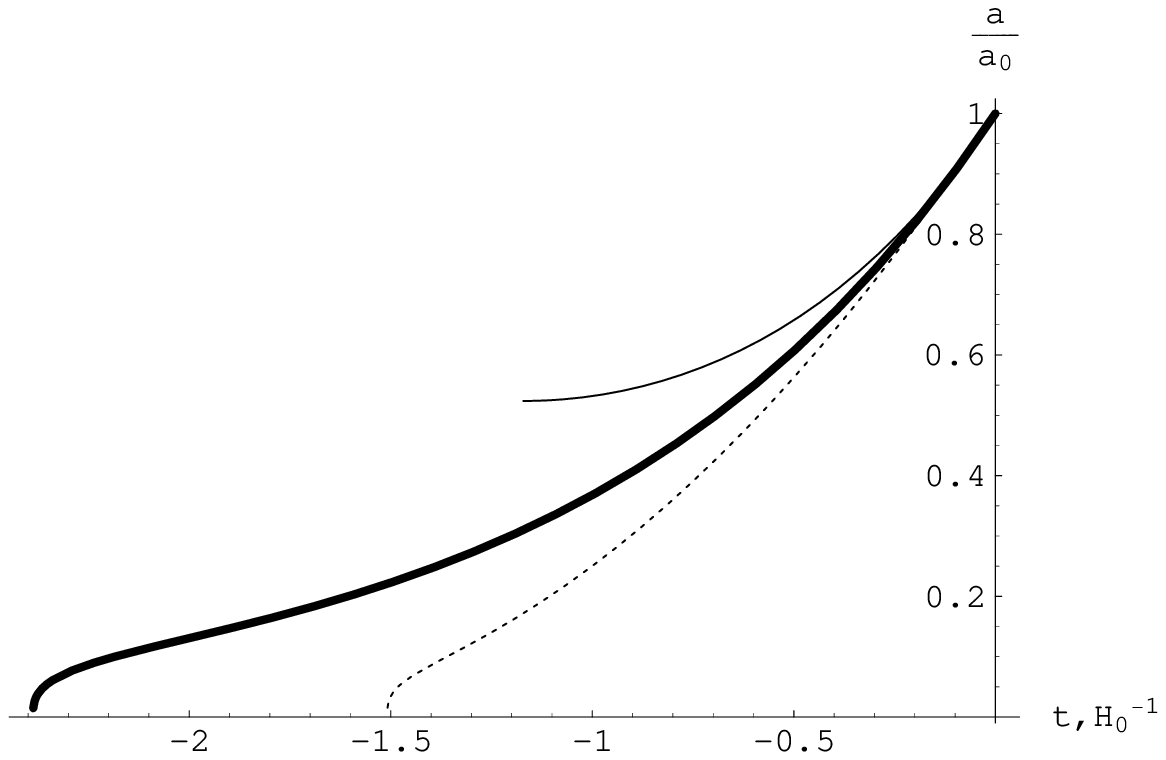}
	\caption{Scale factor for model 2 with radiation as a function of time.}
	\label{fig5}
\end{figure}

Solutions of these equations are shown at fig. \ref{fig3}, \ref{fig4} and \ref{fig5}.

The difference with model 1 is that now energy density and matter density
behave differently (see fig. \ref{fig3} and \ref{fig4}) since in 
the simple relation $\rho =\mu c^{2}$ the speed of light is no more
a constant but a function of the scale factor, so now $\rho \propto \mu a^{2}$.
Separatrix solution is very special since the total energy density for this
solution begins in infinity, decreases, then vanishes at some value of the scale factor, and
increases without bound again for large $a$. For smaller density parameter, than
the separatrix one, solutions admit negative energy density for finite
interval of scale factors, and for larger density parameter the energy
density is always positive definite. In contrast (as can be seen at fig. \ref{fig4}), matter density is always
negative for small scale factors (RD stage). For separatrix solution the
matter density stays constant during "vacuum-dominated" (VD) (dominance of
the vacuum energy density in cosmological equations) epoch, which follows
after MD stage; there matter density decreases for $\Omega _{m}>\Omega _{sep}$ and
increases otherwise.

\subsection{Model 3, with radiation}

Here gravitational constant decreases with cosmic expansion $G\propto a^{-2}$ and it is
assumed that vacuum matter density $\mu _{GX}$ is a new constant.
Cosmological equations read in this case

\begin{eqnarray}
H^{2}+\frac{kc^{2}}{a^{2}} = \frac{\pi ^{2}}{3}\left( \frac{c}{a}\right)
^{2}\left[ 1+\frac{1}{\mu _{GX}}\left( \mu +\frac{\rho _{r0}}{c^{2}}\left( 
\frac{a}{a_{0}}\right) ^{-4}\right) \right] ,   \\
\dot{\mu}+H(\mu -2\mu _{GX}) = 2H\frac{\rho _{r0}}{c^{2}}\left( \frac{a}{%
a_{0}}\right) ^{-4}.
\end{eqnarray}

At fig. 6, 7 and 8 we illustrate solutions of these equations.

\begin{figure}
	\centering
		\includegraphics{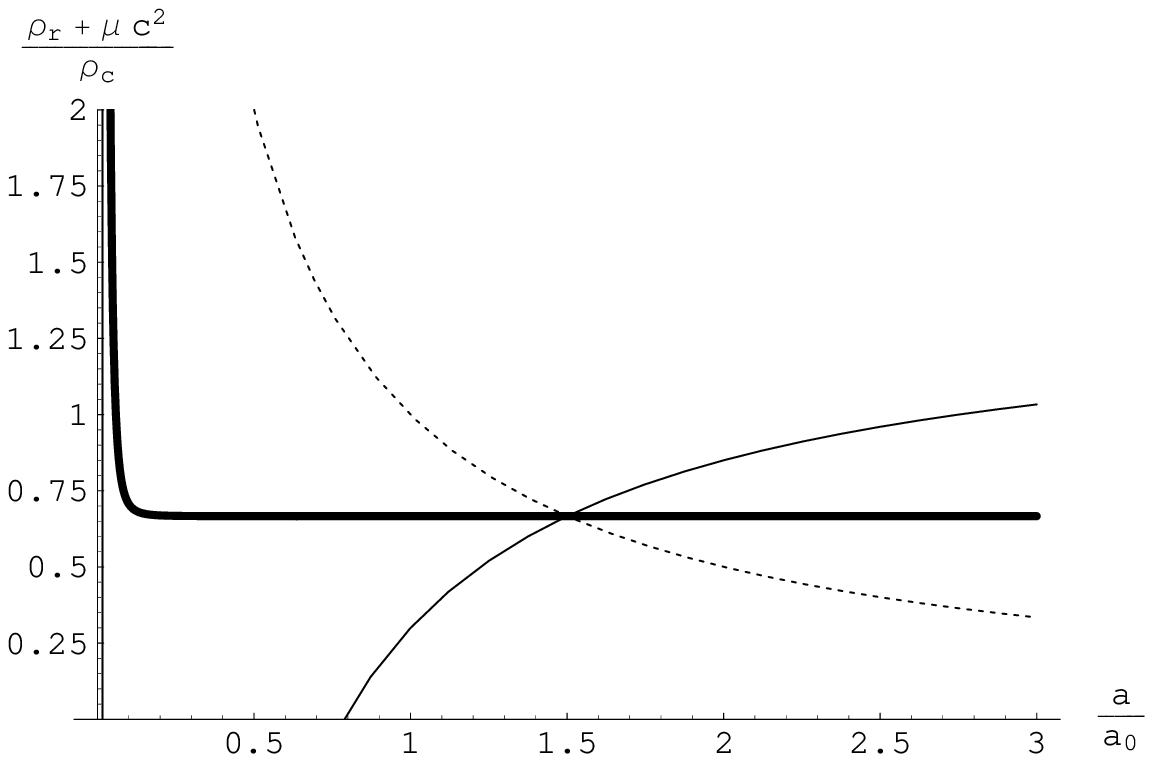}
	\caption{Total energy density as a function of scale factor for model 3 with radiation.}
	\label{fig6}
\end{figure}

\begin{figure}
	\centering
		\includegraphics{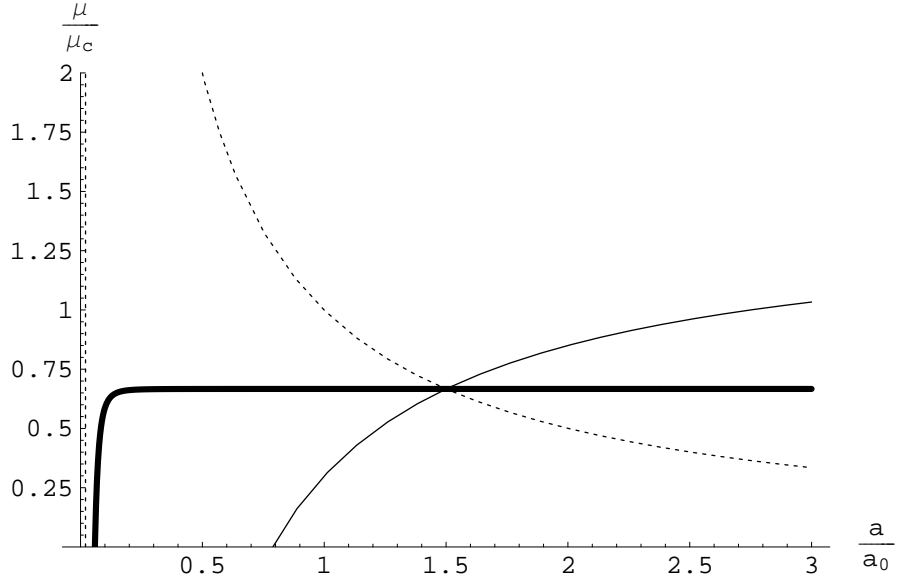}
	\caption{Matter density as a function of scale factor for model 3 with radiation.}
	\label{fig7}
\end{figure}

\begin{figure}
	\centering
		\includegraphics{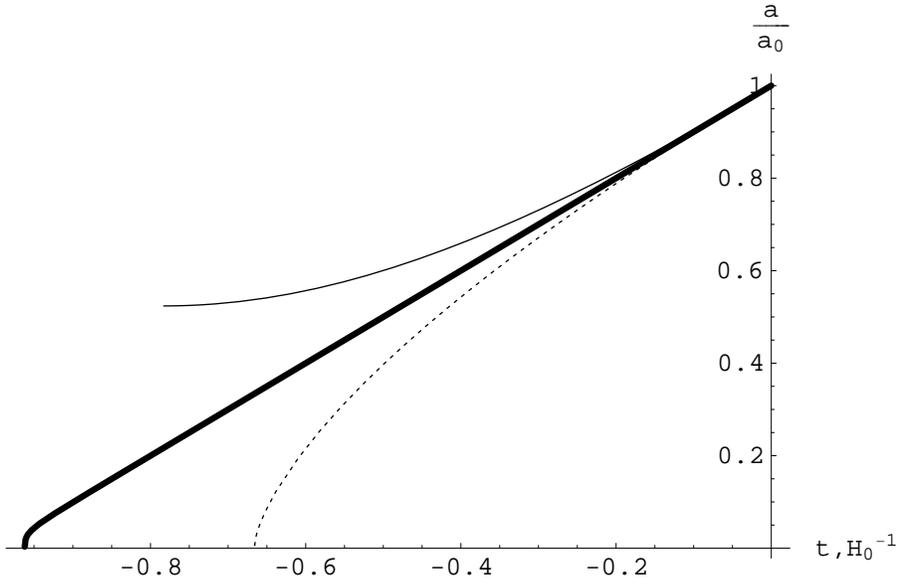}
	\caption{Scale factor for model 3 with radiation.}
	\label{fig8}
\end{figure}

One can see that during the VD stage both total energy density and matter
density of pressureless fluid stay constants for separatrix solution; they
both increase for $\Omega _{m}>\Omega _{sep}$ and decrease otherwise.

\subsection{Model 4, with radiation}

In this model it is assumed that the speed of light changes with time, but
the vacuum energy density $\rho _{GX}$ stays constant. This leads to $%
c\propto a^{1/2}$ dependence.

Cosmological equations are

\begin{eqnarray}
H^{2} = \frac{8\pi G}{3}\left( \mu +\frac{1}{a}\sqrt{\frac{\pi }{8G\rho
_{GX}}}\rho _{r0}\left( \frac{a}{a_{0}}\right) ^{-4}\right) +\frac{\beta }{a}%
,   \\
\dot{\mu}+2H\mu = \frac{H}{a}\sqrt{\frac{\pi \rho _{GX}}{2G}}\left( 1+\frac{%
\rho _{r0}}{\rho _{GX}}\left( \frac{a}{a_{0}}\right) ^{-4}\right) ,
\end{eqnarray}%
where $\beta =\frac{2\sqrt{2}\pi ^{3/2}}{3}\left( G\rho _{GX}\right)
^{1/2}\left( 1-\frac{3k}{\pi ^{2}}\right) $.

Solutions in this case are represented at fig. 9, 10 and 11.

\begin{figure}
	\centering
		\includegraphics{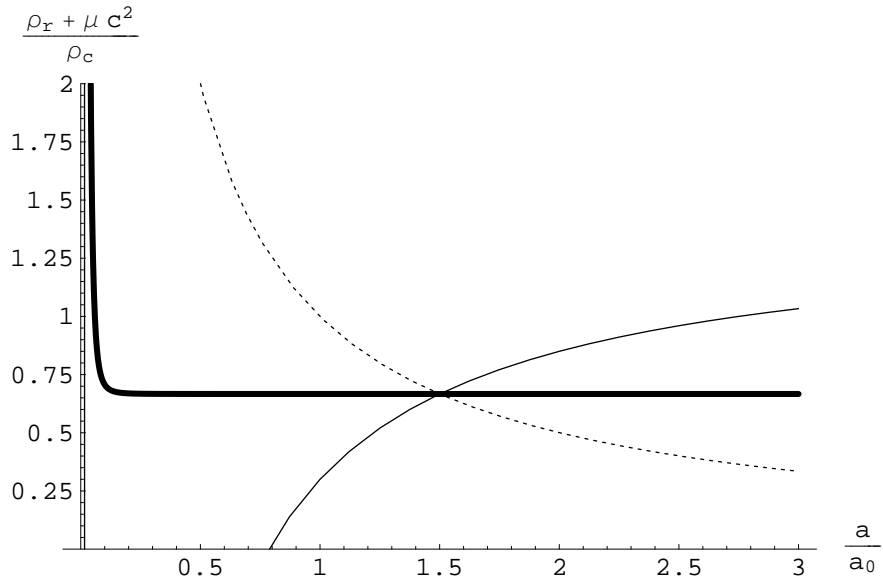}
	\caption{Total energy density in the model 4 with radiation.}
	\label{fig9}
\end{figure}

\begin{figure}
	\centering
		\includegraphics{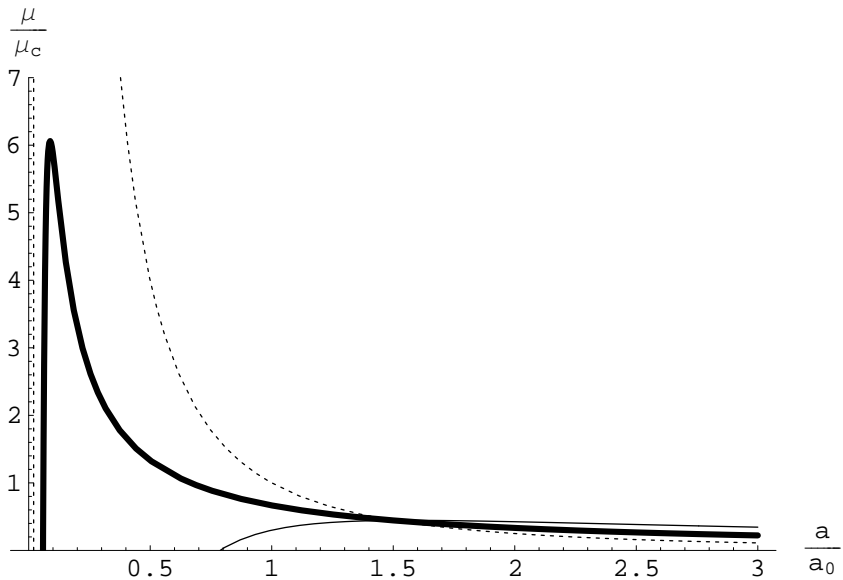}
	\caption{Matter density in the model 4 with radiation.}
	\label{fig10}
\end{figure}

\begin{figure}
	\centering
		\includegraphics{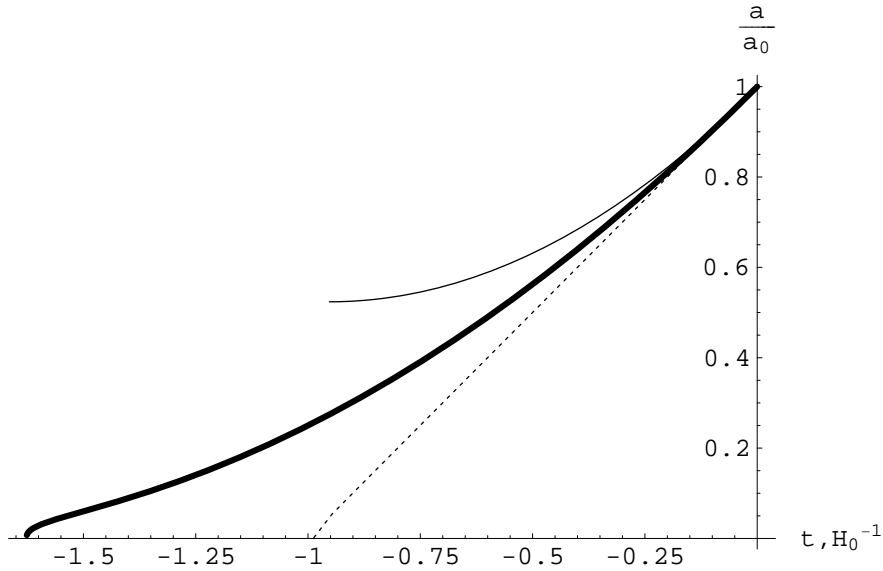}
	\caption{Scale factor in the model 4 with radiation.}
	\label{fig11}
\end{figure}

The role of separatrix in this model is similar to the model 3 with respect
to the energy density, but the matter density of pressureless matter behaves
differently since now $\rho\propto a\mu$.

\section{Cosmological perturbations with varying constants in Newtonian
approximation}

Since cosmological equations and their solutions for GX models are known, it is of interest to
consider evolution of perturbations in these models. This is relevant for
understanding the dynamics of large scale structure, its analogy and
difference with the standard cosmological model. Due to some peculiarities
of GX models this dynamics turns out to be quite different.

We remind that Newtonian equations for perturbations may be used to describe
evolution of perturbations in nonrelativistic component and for modes well
inside horizon.

In general, when both speed of light and gravitational constant change with
time we have (see Appendix):

continuity equation
\begin{equation}
\frac{\partial \mu }{\partial t}+\partial _{\alpha }\left( \mu v^{\alpha
}\right) =-\mu \left( \frac{\dot{G}}{G}-2\frac{\dot{c}}{c}\right) ,
\label{ro1}
\end{equation}%

Euler equation%
\begin{equation}
\frac{\partial \mathbf{v}_{\mathbf{\alpha }}}{\partial t}+\mathbf{v}_{%
\mathbf{\beta }}\frac{\partial \mathbf{v}_{\mathbf{\alpha }}}{\partial
x^{\beta }}+\frac{\partial p}{\mu \partial x^{\alpha }}+\frac{\partial
\varphi }{\partial x^{\alpha }}=-\mathbf{v}_{\mathbf{\alpha }}\frac{\dot{c}}{%
c},
\end{equation}%

and Poisson equation
\begin{equation}
\triangle\varphi -4\pi G\mu =0,  \label{phi1}
\end{equation}%
where $\mu $ and $p$ are density and pressure of the perfect fluid, which is
moving with velocity $\mathbf{v}_{\mathbf{\alpha }}$ in a gravitational
potential $\varphi $. Both the continuity and Euler equations contain
additional terms: $\mu \left( \frac{\dot{G}}{G}-2%
\frac{\dot{c}}{c}\right) $ in continuity equation and $-\mathbf{v}_{\mathbf{%
\alpha }}\frac{\dot{c}}{c}$ term in Euler equation. This means, that mass in
comoving frame is not conserved in models with varying constants, so there
is continuous creation of matter. Also velocity flow changes not only due to
pressure and gravitational field gradients and universal expansion, but also due to variation of the speed of light.

Density perturbations in expanding Universe obey second order differential equation. In  Newtonian approximation, neglecting the sound velocity this equation reads (see Appendix)

\begin{equation*}
\delta ^{^{\prime \prime }}(a)+\delta ^{^{\prime }}(a)\left( \frac{3}{a}+%
\frac{1}{Ha}\frac{\dot{c}}{c}+\frac{H^{^{\prime }}}{H}\right) -\frac{4\pi G%
\bar{\mu}}{H^{2}a^{2}}\delta =0.
\end{equation*}

where $\delta=(\mu-\bar\mu)/\bar\mu$ is dimensionless density contrast, $\bar{\mu}$ is
unperturbed density. Here again, in contrast with the corresponding equation
for Friedmannian models, additional term appears due to variation of the
speed of light. These equations are well defined to describe perturbations in pressureless component which represents e.g. cold dark matter, or baryons after recombination in large wavelength limit.

\section{Linear Newtonian perturbations in GX models}

\subsection{Friedmannian models}

Solutions for density perturbations in Friedmann models are well known.
During RD stage density contrast of dark matter grows\footnote{%
In this paper we consider only perturbations with wavelength larger than the
corresponding Jeans wavelength.} with the scale factor logarithmically%
\footnote{%
We are interested in growing modes only.},

\begin{equation*}
\delta \propto \ln a.
\end{equation*}

During MD epoch perturbations both in baryonic and dark matter components
grow linearly with the scale factor,

\begin{equation*}
\delta \propto a.
\end{equation*}

Finally, in VD, or curvature-dominated (kD) epochs perturbations cease to
grow.

Now we turn to perturbations dynamics specifically within GX models.

\subsection{Model 1 with radiation}

The solution for matter density is 
\begin{equation*}
\bar{\mu}=\frac{c^{2}\pi }{4a^{2}G}\left( 1-\frac{a_{0}}{a}\right) +\left( 
\frac{a}{a_{0}}\right) ^{-3}\mu _{0}
\label{sol1}
\end{equation*}%
and one recovers Friedmannian behavior $\bar{\mu}\propto a^{-3}$ in the
limit of small scale factors. However, it is during RD stage where dynamics
is determined by radiation. According to this solution at late expansion
dominates the curvature.

In this model $c=const,G=const,$ so the perturbation equation is

\begin{equation}
\delta ^{^{\prime \prime }}(a)+\delta ^{^{\prime }}(a)\left( \frac{3}{a}+%
\frac{H^{^{\prime }}}{H}\right) -\frac{4\pi G\bar{\mu}}{H^{2}a^{2}}\delta =0.
\label{main_1}
\end{equation}

\subsubsection{Radiation-dominated epoch.}

Within Newtonian approximation one can still follow dynamics of
perturbations in the pressureless component during RD stage \cite{TSAG}.
This results in the following solution

\begin{equation*}
\delta \propto \ln a,
\end{equation*}%
so perturbations evolve exactly as they do in Friedmannian models. It is not
surprising, since during this stage we have usual dependence of the matter
density on the scale factor, and (\ref{main_1}) is also the same as in
Friedmannian models.

\subsubsection{Matter-dominated epoch.}

Here we take into account only $a^{-3}$ terms in (\ref{sol1}) and also
put $\rho _{r0}\rightarrow 0$, $\acute{k}=0$\ when calculate H(a). The
solution is 
\begin{equation}
\delta \propto a,
\end{equation}%
and it again coincides with the corresponding solution in Friedmannian
models. The reason is the same as for RD stage.

\subsubsection{Curvature dominated epoch.}

Here we account only for $a^{-2}$\ terms in (\ref{sol1}) and also put $%
\rho _{r0}\rightarrow 0$\ when calculate the Hubble rate.

Thus we get the solution

\begin{equation}
\delta \propto a^{\frac{1}{2}\left( \sqrt{5+\frac{4k}{\pi ^{2}-k}}-1\right)
.}
\end{equation}%
\bigskip

Interesting, that for zero spatial curvature this is "golden section" solution

\begin{equation}
\delta \propto a^{\frac{1}{2}\left( \sqrt{5}-1\right) }
\end{equation}

Perturbations grow at kD stage due to existence of energy transfer from dark energy
into pressureless matter which manifests itself in the presence of the
source in continuity equation of this model.

As we see, in contrast with the Friedmann models, perturbations grow during entire cosmological expansion.

\subsection{Model 2, with radiation}

Matter density depends on the scale factor as follows%
\begin{equation}
\bar{\mu}=\frac{\Lambda }{4\pi G}\left( 1-\frac{a_{0}}{a}\right) +\frac{a_{0}%
}{a}\mu _{0}+\frac{4\pi ^{2}\rho _{r0}}{5a_{0}\Lambda }\left[ 1-\left( \frac{%
a_{0}}{a}\right) ^{5}\right]
\label{sol2}
\end{equation}

As in this model $c\propto a,$ the perturbation equation becomes

\begin{equation}
\delta ^{^{\prime \prime }}(a)+\delta ^{^{\prime }}(a)\left( \frac{4}{a}+%
\frac{H^{^{\prime }}}{H}\right) -\frac{4\pi G\bar{\mu}}{H^{2}a^{2}}\delta =0.
\label{main_2}
\end{equation}

In all GX\ models with radiation, except for model 1, matter density is
negative in the beginning of cosmological equations (see figures above).
This leads to anomalous behavior of perturbations during RD stage and in
what follows we do not consider dynamics of perturbations for RD\ epoch.

\subsubsection{Matter-dominated epoch.}

In contrast to model 1, matter-dominated epoch corresponds to the prevalence
of \textit{\ }$\mathit{a}^{-1}$\ terms\ in (\ref{sol2}). We also put $%
\acute{\Lambda}=0$, $\rho _{r0}\rightarrow 0$\ in the background equations when calculate
the Hubble rate.

The solution is

\begin{equation}
\delta \propto \sqrt{a}.
\end{equation}

Thus perturbations growth is somewhat slower comparing to Friedmannian case in MD epoch.
This is a consequence of matter creation in all space due to variation of
the speed of light with cosmic time.

\subsubsection{Vacuum-dominated epoch.}

Here we take into account only $a^{0}$\ terms in (\ref{sol2}) and
also put $\rho _{r0}\rightarrow 0$\ when calculate the Hubble rate.

In this way we obtain the solution%
\begin{equation}
\delta \propto a^{\frac{1}{2}\left( \sqrt{\frac{13\pi ^{2}-9k}{\pi ^{2}-k}}%
-3\right) }.
\end{equation}

Matter is also created during VD stage but this leads to an opposite effect, comparing to the case discussed above, namely it stimulates the growth of perturbations, like at
curvature-dominated stage in the model 1.

For zero spatial curvature $k=0$\ it follows

\begin{equation}
\delta (a)=c_{1}a^{\frac{\sqrt{13}}{2}-\frac{3}{2}}.
\end{equation}

Again, perturbations grow at all stages of expansion.

\subsection{Model 3, with radiation}

Cosmological equations lead to the following solution for matter density%
\begin{equation}
\bar{\mu}=2\mu _{GX}+\frac{a_{0}}{a}\left( \mu _{0}-2\mu _{GX}\right) +\frac{%
2\rho _{r0}}{3c^{2}}\frac{a_{0}}{a}\left( 1-\left( \frac{a_{0}}{a}\right)
^{3}\right) .
\end{equation}

In this model $c=const,$ $G\propto a^{-2},$ so the perturbation equation is
the same as in model 1.

Solutions for perturbations within this model during MD and VD epochs are exactly the same as in model 1 during MD and kD epochs respectively.

\subsection{Model 4, with radiation}

Matter density depends on the scale factor as follows

\begin{equation}
\bar{\mu}=\left( \frac{a_{0}}{a}\right) ^{2}\mu _{0}+\frac{1}{a}\left( 1-%
\frac{a_{0}}{a}\right) \sqrt{\frac{\pi \rho _{GX}}{2G}}+\frac{\rho _{r0}a_{0}%
}{3}\frac{1}{a^{2}}\sqrt{\frac{\pi }{2G\rho _{GX}}}\left( 1-\left( \frac{%
a_{0}}{a}\right) ^{3}\right) .
\label{sol4}  
\end{equation}

In this model $c\propto a^{\frac{1}{2}},$ $G=const,$ so the perturbation
equation is modified

\begin{equation}
\delta ^{^{\prime \prime }}(a)+\delta ^{^{\prime }}(a)\left( \frac{7}{2a}+%
\frac{H^{^{\prime }}}{H}\right) -\frac{4\pi G\bar{\mu}}{H^{2}a^{2}}\delta =0.
\label{main_4}
\end{equation}

\subsubsection{Matter-dominated epoch.}

Considering only $a^{-2}$ terms in (\ref{sol4}) and assuming in addition $%
\beta =0$, $\rho _{r0}\rightarrow 0$ in the cosmological equations we find the solution

\begin{equation}
\delta \propto a^{\frac{\sqrt{33}}{4}-\frac{3}{4}}.
\end{equation}

This is slower growth than in the Friedmannian model and model 1, but faster
than in the model 2.

\subsubsection{$\protect\beta $-dominated epoch.}

Taking into account only $a^{-1}$\ terms in (\ref{sol4}) and also putting $%
\rho _{r0}\rightarrow 0$\ to calculate the Hubble rate we find the solution

\begin{equation}
\delta \propto a^{\sqrt{\frac{2\pi ^{2}-k}{\pi ^{2}-k}}-1}.
\end{equation}

For zero spatial curvature we obtain

\begin{equation}
\delta (a)=a^{\sqrt{2}-1}.
\end{equation}

Again, perturbations grow during entire expansion.

\section{Conclusions}

Dynamics of perturbations in Gurzadyan-Xue cosmological models is studied within Newtonian approximation. With this goal relativistic hydrodynamic equations for models with varying constants are derived. Non-relativistic limit for these equations is also obtained, which provided the basis for analysis of perturbations dynamics within Newtonian approximation. Analytic solutions are presented for the model with pressureless ideal fluid and radiation. In contrast with the standard cosmological model perturbations grow during entire cosmological expansion, including curvature-dominated and vacuum-dominated epochs.

Only Poisson equation preserves its form comparing to the Newtonian hydrodynamics. Both continuity and Euler equations contain additional terms due to variation of the physical constants, namely the speed of light and the gravitational constant. Thus both at cosmological and at local level continuity equation contains source terms which may be interpreted as creation of matter.

We found no difference in dynamics of perturbations between our model 1 (varying vacuum energy density) and standard cosmological model for radiation- and matter-dominated stages. However, on the last stage perturbations grow according to the ``golden-section'' rule. In other models we have found polynomial dependence of the density contrast on the scale factor.

\section{Appendix}

\subsection{Hydrodynamic equations}

It is our goal in this section to derive hydrodynamic equations for models
with varying speed of light and gravitational constant in Newtonian limit.
We start with the interval

\begin{equation*}
ds^{2}=\left( c^{2}+2\varphi \right) dt^{2}-dr_{\alpha }^{2},
\end{equation*}

where $r_{\alpha }=\left( x,y,z\right) $ are cartesian coordinates, so
metric coefficients are

\begin{equation*}
g_{00}=c^{2}+2\varphi ,\ \ \ g_{11}=g_{22}=g_{33}=-1.
\end{equation*}

In Newtonian limit

\begin{equation*}
\frac{2\varphi }{c^{2}}\ll 1
\end{equation*}

is a small parameter, and in Taylor expansion we will keep only zeroth order
terms with respect to it.

Christoffel symbol has the following nonvanishing components

\begin{eqnarray*}
\Gamma _{00}^{0} &=&\frac{c\dot{c}+\dot{\varphi}}{c^{2}+2\varphi },\ \ \
\Gamma _{00}^{\alpha }=\frac{\partial \varphi }{\partial x^{\alpha }}, \\
\Gamma _{\alpha 0}^{0} &=&\frac{1}{c^{2}+2\varphi }\frac{\partial \varphi }{%
\partial x^{\alpha }}.
\end{eqnarray*}

Here Latin indexes take values from 0 to 3, while Greek indexes run from 1
to 3; a dot denotes differentiation with respect to time.

Ricci tensor has the following nonvanishing components

\begin{eqnarray*}
R_{\beta }^{\alpha } &=&\delta _{\alpha \beta }\frac{1}{c^{2}+2\varphi }%
\left[ \frac{\partial ^{2}\varphi }{\partial \left( x^{\alpha }\right) ^{2}}-%
\frac{1}{c^{2}+2\varphi }\left( \frac{\partial \varphi }{\partial x^{\alpha }%
}\right) ^{2}\right] , \\
R_{0}^{0} &=&\frac{1}{c^{2}+2\varphi }\left[ \Delta \varphi -\frac{1}{%
c^{2}+2\varphi }\left( \nabla \varphi \right) ^{2}\right] \approx \frac{1}{%
c^{2}}\Delta \varphi .
\end{eqnarray*}

\subsubsection{Poisson equation.}

From Einstein's equations

\begin{equation*}
R_{k}^{i}=\kappa \left( T_{k}^{i}-\frac{1}{2}g_{k}^{i}T\right) ,
\end{equation*}

where

\begin{equation*}
\kappa =\frac{8\pi G}{c^{4}},
\end{equation*}

for the fluid at rest one has

\begin{equation*}
R_{0}^{0}=\frac{4\pi G}{c^{2}}\mu ,
\end{equation*}

so we obtain the \emph{usual Poisson equation}

\begin{equation*}
\Delta \varphi =4\pi G\mu .
\end{equation*}

Energy-momentum tensor for a perfect fluid is

\begin{equation*}
T_{i}\ ^{k}=\omega u^{k}u_{i}-pg_{i}\ ^{k},
\end{equation*}

where $\omega =\mu c^{2}+p,$ $u^{i}=\frac{\gamma }{\sqrt{c^{2}+2\varphi }}(1,%
\mathbf{v}_{\mathbf{\alpha }}),$ $u_{i}=\frac{\gamma }{\sqrt{c^{2}+2\varphi }%
}(c^{2}+2\varphi ,-\mathbf{v}_{\mathbf{\alpha }}),$ $\gamma =\left( 1-\frac{%
\mathbf{v}^{2}}{c^{2}}\right) ^{-\frac{1}{2}}$.

In components

\begin{eqnarray*}
T_{0}\ ^{0} &=&\omega \gamma ^{2}-p, \\
T_{0}\ ^{\mathbf{\alpha }} &=&\omega \gamma ^{2}\mathbf{v}_{\mathbf{\alpha }%
}=-c^{2}T_{\mathbf{\alpha }}\ ^{0}, \\
T_{\alpha }\ ^{\mathbf{\beta }} &=&-\omega \left( \frac{\gamma }{c}\right)
^{2}\mathbf{v}_{\mathbf{\alpha }}\mathbf{v}_{\mathbf{\beta }}-p\delta
_{\alpha \beta }.
\end{eqnarray*}

Euler and continuity equations both follow from energy-momentum conservation

\begin{equation*}
\frac{1}{\kappa }\left( \kappa T_{i}\ ^{k}\right) _{;k}=\frac{1}{\sqrt{-g}}%
\frac{\partial \sqrt{-g}}{\partial x^{k}}T_{i}\ ^{k}+\frac{\dot{\kappa}}{%
\kappa }T_{i}\ ^{0}+\frac{\partial }{\partial x^{k}}\left( T_{i}\
^{k}\right) -\frac{1}{2}\frac{\partial g_{kl}}{\partial x^{i}}T^{kl}=0.
\end{equation*}

We again expand all terms in this equation in Taylor series with respect to
small parameters

\begin{eqnarray*}
\frac{\mathbf{v}_{\mathbf{\alpha }}}{c} &\ll &1, \\
\frac{p}{\mu c^{2}} &\ll &1.
\end{eqnarray*}

\subsubsection{Continuity equation.}

From zeroth component of energy-momentum conservation we find

\begin{displaymath}
\begin{array}{lcr}
\displaystyle{\frac{1}{\sqrt{-g}}\frac{\partial \sqrt{-g}}{\partial x^{k}}T_{0}\ ^{k}+%
\frac{\dot{\kappa}}{\kappa }T_{0}\ ^{0}+\frac{\partial }{\partial x^{k}}%
\left( T_{0}\ ^{k}\right) -\frac{1}{2}\frac{\partial g_{kl}}{\partial t}%
T^{kl}=} \\
\displaystyle{=\frac{1}{\sqrt{-g}}\left( \frac{\partial \sqrt{-g}}{\partial t}T_{0}\ ^{0}+%
\frac{\partial \sqrt{-g}}{\partial x^{\alpha }}T_{0}\ ^{\alpha }\right) +%
\frac{\dot{\kappa}}{\kappa }T_{0}\ ^{0}+\frac{\partial }{\partial t}\left(
T_{0}\ ^{0}\right) +\frac{\partial }{\partial x^{\alpha }}\left( T_{0}\
^{\alpha }\right)-} \\
\displaystyle{-\Gamma _{00}^{0}T_{0}\ ^{0}=\Gamma _{00}^{0}T_{0}\ ^{0}+\Gamma _{\alpha 0}^{0}T_{0}\ ^{\alpha }+\frac{%
\dot{\kappa}}{\kappa }T_{0}\ ^{0}+\frac{\partial }{\partial t}\left( T_{0}\
^{0}\right) +\frac{\partial }{\partial x^{\alpha }}\left( T_{0}\ ^{\alpha
}\right) -} \\
\displaystyle{-\Gamma _{00}^{0}T_{0}\ ^{0}=\frac{1}{c^{2}+2\varphi }\frac{\partial \varphi }{\partial x^{\alpha }}%
\omega \gamma ^{2}\mathbf{v}_{\mathbf{\alpha }}+\left( \frac{\dot{G}}{G}-4%
\frac{\dot{c}}{c}\right) \left( \omega \gamma ^{2}-p\right) +\frac{\partial 
}{\partial t}\left( \omega \gamma ^{2}-p\right)+} \\
\displaystyle{+\frac{\partial }{\partial
x^{\alpha }}\left( \omega \gamma ^{2}\mathbf{v}_{\mathbf{\alpha }}\right)
\approx\frac{1}{c^{2}}\frac{\partial \varphi }{\partial x^{\alpha }}\omega
\gamma ^{2}\mathbf{v}_{\mathbf{\alpha }}+\left( \frac{\dot{G}}{G}-4\frac{%
\dot{c}}{c}\right) \left( \omega \gamma ^{2}-p\right) +\frac{\partial }{%
\partial t}\left( \omega \gamma ^{2}-p\right)+} \\
\displaystyle{+\frac{\partial }{\partial
x^{\alpha }}\left( \omega \gamma ^{2}\mathbf{v}_{\mathbf{\alpha }}\right) =0.}
\end{array}
\end{displaymath}

Taking Newtonian limit of this expression, and dividing all terms by $c^{2}$
we find

\begin{displaymath}
\begin{array}{lcr}
\displaystyle{\frac{1}{c^{2}}\frac{\partial \varphi }{\partial x^{\alpha }}\mu \mathbf{v}_{%
\mathbf{\alpha }}+\frac{1}{c^{2}}\left( \frac{\dot{G}}{G}-4\frac{\dot{c}}{c}%
\right) \mu c^{2}+\frac{1}{c^{2}}\frac{\partial }{\partial t}\left( \mu
c^{2}\right) +\frac{1}{c^{2}}\frac{\partial }{\partial x^{\alpha }}\left(
\mu c^{2}\mathbf{v}_{\mathbf{\alpha }}\right) =} \\
\displaystyle{\frac{1}{c^{2}}\frac{\partial \varphi }{\partial x^{\alpha }}\mu \mathbf{v}_{%
\mathbf{\alpha }}+\left( \frac{\dot{G}}{G}-4\frac{\dot{c}}{c}\right) \mu
+2\mu \frac{\dot{c}}{c}+\frac{\partial \mu }{\partial t}+\frac{\partial }{%
\partial x^{\alpha }}\left( \mu \mathbf{v}_{\mathbf{\alpha }}\right) =} \\
\displaystyle{\frac{\partial \mu }{\partial t}+\frac{\partial }{\partial x^{\alpha }}%
\left( \mu \mathbf{v}_{\mathbf{\alpha }}\right) +\left( \frac{\dot{G}}{G}-2%
\frac{\dot{c}}{c}\right) \mu +\frac{1}{c^{2}}\frac{\partial \varphi }{%
\partial x^{\alpha }}\mu \mathbf{v}_{\mathbf{\alpha }}=0}.
\end{array}
\end{displaymath}

The last term is small comparing to the second, as can be seen from the
following

\begin{displaymath}
\begin{array}{lcr}
\displaystyle{2\varphi \ll c^{2},} \\
\displaystyle{2\mu \mathbf{v}_{\mathbf{\alpha }}\varphi \ll \mu \mathbf{v}_{\mathbf{\alpha 
}}c^{2},} \\
\displaystyle{2\frac{\partial }{\partial x^{\alpha }}\left( \mu \mathbf{v}_{\mathbf{\alpha 
}}\varphi \right) \ll c^{2}\frac{\partial \left( \mu \mathbf{v}_{\mathbf{%
\alpha }}\right) }{\partial x^{\alpha }},} \\
\displaystyle{2\mu \mathbf{v}_{\mathbf{\alpha }}\frac{\partial \varphi }{\partial
x^{\alpha }}+2\varphi \frac{\partial \left( \mu \mathbf{v}_{\mathbf{\alpha }%
}\right) }{\partial x^{\alpha }}\ll c^{2}\frac{\partial \left( \mu \mathbf{v}%
_{\mathbf{\alpha }}\right) }{\partial x^{\alpha }}}.
\end{array}
\end{displaymath}

Therefore, \emph{continuity equation} for models with varying constants in
Newtonian approximation reads

\begin{equation*}
\frac{\partial \mu }{\partial t}+\frac{\partial }{\partial x^{\alpha }}%
\left( \mu \mathbf{v}_{\mathbf{\alpha }}\right) =\left( 2\frac{\dot c}{c}-%
\frac{\dot G}{G}\right) \mu .
\end{equation*}

\subsubsection{Euler equation.}

For the remaining 3 components of the energy-momentum tensor we have%
\begin{displaymath}
\begin{array}{lcr}
\displaystyle{\frac{1}{\sqrt{-g}}\frac{\partial \sqrt{-g}}{\partial x^{k}}T_{\alpha }\
^{k}+\frac{\dot{\kappa}}{\kappa }T_{\alpha }\ ^{0}+\frac{\partial }{\partial
x^{k}}\left( T_{\alpha }\ ^{k}\right) -\frac{1}{2}\frac{\partial g_{kl}}{%
\partial x^{\alpha }}T^{kl}=} \\
\displaystyle{=\frac{1}{\sqrt{-g}}\left( \frac{\partial \sqrt{-g}}{\partial t}T_{\alpha }\
^{0}+\frac{\partial \sqrt{-g}}{\partial x^{\beta }}T_{\alpha }\ ^{\beta
}\right) +\frac{\dot{\kappa}}{\kappa }T_{\alpha }\ ^{0}+\frac{\partial }{%
\partial t}\left( T_{\alpha }\ ^{0}\right) +} \\
\displaystyle{+\frac{\partial }{\partial
x^{\beta }}\left( T_{\alpha }\ ^{\beta }\right) -\frac{1}{2}\frac{\partial
g_{00}}{\partial x^{\alpha }}T^{00}=\Gamma _{00}^{0}T_{\alpha }\ ^{0}+\Gamma _{\beta 0}^{0}T_{\alpha }\ ^{\beta
}+\frac{\dot{\kappa}}{\kappa }T_{\alpha }\ ^{0}+\frac{\partial }{\partial t}%
\left( T_{\alpha }\ ^{0}\right) +} \\ 
\displaystyle{+\frac{\partial }{\partial x^{\beta }}\left(
T_{\alpha }\ ^{\beta }\right) -\Gamma _{\alpha 0}^{0}T_{0}\ ^{0}=-\frac{c\dot{c}+\dot{\varphi}}{c^{2}+2\varphi }\omega \left( \frac{\gamma }{%
c}\right) ^{2}\mathbf{v}_{\mathbf{\alpha }}-} \\
\displaystyle{-\frac{1}{c^{2}+2\varphi }\frac{%
\partial \varphi }{\partial x^{\beta }}\left[ \omega \left( \frac{\gamma }{c}%
\right) ^{2}\mathbf{v}_{\mathbf{\alpha }}\mathbf{v}_{\mathbf{\beta }%
}+p\delta _{\alpha \beta }\right] -\left( \frac{\dot{G}}{G}-4\frac{\dot{c}}{c%
}\right) \omega \left( \frac{\gamma }{c}\right) ^{2}\mathbf{v}_{\mathbf{%
\alpha }}-} \\ 
\displaystyle{-\frac{\partial }{\partial t}\left[ \omega \left( \frac{\gamma }{c}\right)
^{2}\mathbf{v}_{\mathbf{\alpha }}\right] -\frac{\partial }{\partial x^{\beta
}}\left[ \omega \left( \frac{\gamma }{c}\right) ^{2}\mathbf{v}_{\mathbf{%
\alpha }}\mathbf{v}_{\mathbf{\beta }}+p\delta _{\alpha \beta }\right] +\frac{%
1}{c^{2}+2\varphi }\frac{\partial \varphi }{\partial x^{\alpha }}\left(
\omega \gamma ^{2}-p\right) \approx } \\
\displaystyle{\approx-\frac{\dot{c}}{c}\omega \left( \frac{\gamma }{c}\right) ^{2}\mathbf{%
v}_{\mathbf{\alpha }}-\frac{1}{c^{2}}\frac{\partial \varphi }{\partial
x^{\beta }}\omega \left( \frac{\gamma }{c}\right) ^{2}\mathbf{v}_{\mathbf{%
\alpha }}\mathbf{v}_{\mathbf{\beta }}-\left( \frac{\dot{G}}{G}-4\frac{\dot{c}%
}{c}\right) \omega \left( \frac{\gamma }{c}\right) ^{2}\mathbf{v}_{\mathbf{%
\alpha }}-}\\
\displaystyle{-\frac{\partial }{\partial t}\left[ \omega \left( \frac{\gamma }{c}\right)
^{2}\mathbf{v}_{\mathbf{\alpha }}\right] -\frac{\partial }{\partial x^{\beta
}}\left[ \omega \left( \frac{\gamma }{c}\right) ^{2}\mathbf{v}_{\mathbf{%
\alpha }}\mathbf{v}_{\mathbf{\beta }}+p\delta _{\alpha \beta }\right] +\frac{%
1}{c^{2}}\frac{\partial \varphi }{\partial x^{\alpha }}\left( \omega \gamma
^{2}-2p\right) =0.}
\end{array}
\end{displaymath}

Taking Newtonian limit of this expression we get%
\begin{displaymath}
\begin{array}{lcr}
\displaystyle{-\frac{\dot{c}}{c}\mu \mathbf{v}_{\mathbf{\alpha }}-\mu \frac{\partial
\varphi }{\partial x^{\beta }}\frac{\mathbf{v}_{\mathbf{\alpha }}\mathbf{v}_{%
\mathbf{\beta }}}{c^{2}}-\left( \frac{\dot{G}}{G}-4\frac{\dot{c}}{c}\right)
\mu \mathbf{v}_{\mathbf{\alpha }}-\frac{\partial }{\partial t}\left( \mu 
\mathbf{v}_{\mathbf{\alpha }}\right) -\frac{\partial }{\partial x^{\beta }}%
\left( \mu \mathbf{v}_{\mathbf{\alpha }}\mathbf{v}_{\mathbf{\beta }}\right)-} \\
\displaystyle{ -\frac{\partial p}{\partial x^{\alpha }}-\mu \frac{\partial \varphi }{%
\partial x^{\alpha }}=-\frac{\partial }{\partial x^{\beta }}\left( \mu \mathbf{v}_{\mathbf{\alpha }}\mathbf{v}_{\mathbf{\beta }}\right) -\frac{\partial p}{\partial x^{\alpha }}%
-\frac{\partial \left( \mu \mathbf{v}_{\mathbf{\alpha }}\right) }{\partial t}%
-\mu \mathbf{v}\left( \frac{\dot{G}}{G}-3\frac{\dot{c}}{c}\right)-} \\
\displaystyle{ -\mu \frac{\partial \varphi }{\partial x^{\alpha }}=-\mathbf{v}_{\mathbf{\alpha }}\left[ \mu \left( \frac{\dot{G}}{G}-2\frac{%
\dot{c}}{c}\right) +\frac{\partial \mu }{\partial t}+\frac{\partial }{%
\partial x^{\beta }}\left( \mu \mathbf{v}^{\beta }\right) \right] +\mu 
\mathbf{v}_{\mathbf{\alpha }}\frac{\dot{c}}{c}-\mu \frac{\partial \mathbf{v}%
_{\mathbf{\alpha }}}{\partial t}-}\\
\displaystyle{-\mu \mathbf{v}_{\mathbf{\beta }}\frac{%
\partial \mathbf{v}_{\mathbf{\alpha }}}{\partial x^{\beta }}-\frac{\partial p%
}{\partial x^{\alpha }}-\mu \frac{\partial \varphi }{\partial x^{\alpha }}=0.}
\end{array}
\end{displaymath}

Finally, dividing this by $\mu $\ we find the \emph{Euler equation} for
models with varying constants in Newtonian approximation

\begin{equation*}
\frac{\partial \mathbf{v}_{\mathbf{\alpha }}}{\partial t}+\mathbf{v}_{%
\mathbf{\beta }}\frac{\partial \mathbf{v}_{\mathbf{\alpha }}}{\partial
x^{\beta }}+\frac{\partial p}{\mu \partial x^{\alpha }}+\frac{\partial
\varphi }{\partial x^{\alpha }}=-\mathbf{v}_{\mathbf{\alpha }}\frac{\dot{c}}{%
c}.
\end{equation*}

\subsection{Perturbation equations}

First, introduce comoving coordinates,%
\begin{equation*}
r_{\alpha }=ax_{\alpha },
\end{equation*}%
then for velocity we have

\begin{equation*}
\mathbf{v}_{\mathbf{\alpha }}=Hr_{\alpha }+u_{\alpha }.
\end{equation*}

Perturbation equations now read

\begin{equation}
\frac{\partial \mu }{\partial t}+3H\mu +\frac{1}{a}\partial _{\alpha }\left(
\mu u_{\alpha }\right) +\mu \left( \frac{\dot{G}}{G}-2\frac{\dot{c}}{c}%
\right) =0,  \label{1}
\end{equation}%
\begin{equation}
\frac{d^{2}a}{dt^{2}}x_{\alpha }+\frac{\partial u_{\alpha }}{\partial t}%
+Hu_{\alpha }+\frac{1}{a}u^{\beta }\partial _{\beta }u_{\alpha }+\frac{1}{%
a\mu }\partial _{\alpha }p+\frac{1}{a}\partial _{\alpha }\varphi +\left(
Hax_{\alpha }+u_{\alpha }\right) \frac{\dot{c}}{c}=0,  \label{2}
\end{equation}

\begin{equation}
\partial ^{2}\varphi -4\pi Ga^{2}\mu =0,  \label{3}
\end{equation}%
where $\partial _{\alpha }=\frac{\partial }{\partial x_{\alpha }},$ $%
\partial ^{2}=\frac{\partial ^{2}}{\partial x_{\alpha }^{2}}$. Differential
operators transform from physical to comoving coordinates as $\frac{\partial 
}{\partial r_{\alpha }}\rightarrow \frac{1}{a}\partial _{\alpha }$, $\frac{%
\partial }{\partial t}\rightarrow \frac{\partial }{\partial t}-Hx_{\beta
}\partial _{\beta }$.

Background solution for $\mu =\bar{\mu}(t)$ , $p=\bar{p}(t)$,$\ \varphi =%
\bar{\varphi}(t)$ and $u_{\alpha }=0$ is given by the system

\begin{eqnarray*}
\frac{\partial \bar{\mu}}{\partial t}+3H\bar{\mu} = -\bar{\mu}\left( \frac{%
\dot{G}}{G}-2\frac{\dot{c}}{c}\right) , \\
\frac{d^{2}a}{dt^{2}}x_{\alpha }+\frac{1}{a}\partial _{\alpha }\bar{\varphi}
= -Hax_{\alpha }\frac{dc}{cdt}, \\
\partial ^{2}\bar{\varphi}-4\pi Ga^{2}\bar{\mu} = 0,
\end{eqnarray*}

Consider perturbations about the background solution

\begin{eqnarray}
\mu &=&\bar{\mu}(1+\delta ),\ \ \ p=\bar{p}(1+\delta p),\ \ \ \varphi =\bar{%
\varphi}(1+\delta \varphi ),\ \ \ u_{\alpha }=\delta u_{\alpha },  \label{5}
\\
\delta &\ll &1,\ \ \ \delta p\ll 1,\ \ \ \delta \varphi \ll 1,\ \ \ \delta
u_{\alpha }\ll 1.
\end{eqnarray}%
From (\ref{1}) and (\ref{5}) we have%
\begin{displaymath}
\begin{array}{lcr}
\displaystyle{\frac{\partial }{\partial t}\left[ \bar{\mu}\left( 1+\delta \mu \right) %
\right] +3H\left[ \bar{\mu}\left( 1+\delta \mu \right) \right] +\frac{1}{a}%
\partial _{\alpha }\left\{ \left[ \bar{\mu}\left( 1+\delta \mu \right) %
\right] \delta u_{\alpha }\right\} =} \\
\displaystyle{=-\left( \frac{dG}{Gdt}-2\frac{dc}{cdt}%
\right) \bar{\mu}\left( 1+\delta \mu \right)}\\
\displaystyle{\delta \frac{\partial \bar{\mu}}{\partial t}+\bar{\mu}\frac{\partial \delta 
}{\partial t}+3H\bar{\mu}\delta +\frac{\bar{\mu}}{a}\partial _{\alpha
}\left( \delta u_{\alpha }\right) =-f\bar{\mu}\delta} \\
\displaystyle{\delta \left[ \frac{\partial \bar{\mu}}{\partial t}+3H\bar{\mu}+\bar{\mu}%
\left( \frac{\dot{G}}{G}-2\frac{\dot{c}}{c}\right) \right] +\bar{\mu}\left[ 
\frac{\partial \delta }{\partial t}+\frac{1}{a}\partial _{\alpha }\left(
\delta u_{\alpha }\right) \right] =0}.
\end{array}
\end{displaymath}

Then

\begin{equation}
\frac{\partial \delta }{\partial t}+\frac{1}{a}\partial _{\alpha }\left(
\delta u_{\alpha }\right) =0.  \label{11}
\end{equation}%
\bigskip

From (\ref{2}) and (\ref{5}) we obtain%
\begin{displaymath}
\begin{array}{lcr}
\displaystyle{\frac{d^{2}a}{dt^{2}}x_{\alpha }+\frac{\partial \delta u_{\alpha }}{\partial
t}+H\delta u_{\alpha }+\frac{1}{a\bar{\mu}\left( 1+\delta \right) }\partial
_{\alpha }\left[ \bar{p}(1+\delta p)\right] +} \\
\displaystyle{+\frac{1}{a}\partial _{\alpha }\left[ \bar{\varphi}(1+\delta \varphi )\right]
=-\frac{dc}{cdt}\left( Hax_{\alpha }+\delta u_{\alpha }\right) ,} \\
\displaystyle{\frac{\partial \delta u_{\alpha }}{\partial t}+H\delta u_{\alpha }+\frac{%
\bar{p}}{a\bar{\mu}}\partial _{\alpha }p+\frac{1}{a}\partial _{\alpha
}\left( \delta \varphi \right) =-\delta u_{\alpha }\frac{dc}{cdt}}.
\end{array}
\end{displaymath}

Assume, as usual, the speed of sound is $v_{s}^{2}=\frac{\partial p}{%
\partial \mu }$, so $\frac{\bar{p}}{\bar{\mu}}\partial _{\alpha }\left(
\delta p\right) =v_{s}^{2}\partial _{\alpha }\delta $, so

\begin{equation}
\frac{\partial \delta u_{\alpha }}{\partial t}+H\delta u_{\alpha }+\frac{%
v_{s}^{2}}{a}\partial _{\alpha }\delta +\frac{1}{a}\partial _{\alpha }\left(
\delta \varphi \right) =-\delta u_{\alpha }\frac{dc}{cdt}.  \label{22}
\end{equation}

From (\ref{3}) and (\ref{5}) we have

\begin{equation}
\partial ^{2}\delta \varphi -4\pi Ga^{2}\bar{\mu}\delta =0.  \label{33}
\end{equation}%
\bigskip

Combine (\ref{11}),(\ref{22}),(\ref{33})\ to a second order differential
equation%
\begin{displaymath}
\begin{array}{lcr}
\displaystyle{\frac{\partial ^{2}\delta }{\partial t^{2}}+\frac{\partial }{\partial t}%
\left[ \frac{1}{a}\partial _{\alpha }\left( \delta u_{\alpha }\right) \right]
=} \\
\displaystyle{=\frac{\partial ^{2}\delta }{\partial t^{2}}+\frac{H}{a}\partial _{\alpha
}\delta u_{\alpha }+\frac{1}{a}\partial _{\alpha }\left( \frac{\partial
\delta u_{\alpha }}{\partial t}\right) =} \\
\displaystyle{=\frac{\partial ^{2}\delta }{\partial t^{2}}-\frac{H}{a}\partial _{\alpha
}\delta \mathbf{v}_{\mathbf{\alpha }}-\frac{1}{a}\partial _{\alpha }\left(
H\delta u_{\alpha }+\frac{v_{s}^{2}}{a}\partial _{\alpha }\delta +\frac{1}{a}%
\partial _{\alpha }\left( \delta \varphi \right) +\delta u_{\alpha }\frac{dc%
}{cdt}\right) =} \\
\displaystyle{=\frac{\partial ^{2}\delta }{\partial t^{2}}-\left( 2H+\frac{\dot{c}}{c}%
\right) \frac{1}{a}\partial _{\alpha }\delta u_{\alpha }-\frac{v_{s}^{2}}{%
a^{2}}\partial ^{2}\delta -4\pi G\bar{\mu}\delta .}
\end{array}
\end{displaymath}
\bigskip

Finally, equation for density perturbations is

\begin{equation*}
\frac{\partial ^{2}\delta }{\partial t^{2}}+\left( 2H+\frac{\dot{c}}{c}%
\right) \frac{\partial \delta }{\partial t}-\frac{v_{s}^{2}}{a^{2}}\partial
^{2}\delta -4\pi G\bar{\mu}\delta =0.
\end{equation*}

The same equation may be rewritten where all functions depend on scale
factor instead of time with conversion of the derivatives 
\begin{eqnarray}
\dot{\delta} &=&Ha\delta ^{^{\prime }}(a) \\
\ddot{\delta} &=&H^{2}a^{2}\delta ^{^{\prime \prime }}(a)+\left[ H^{^{\prime
}}(a)Ha^{2}+H^{2}a\right] \delta ^{^{\prime }}(a)
\end{eqnarray}

Then density perturbations satisfy the following equation

\begin{equation}
\delta ^{^{\prime \prime }}(a)+\delta ^{^{\prime }}(a)\left( \frac{3}{a}+%
\frac{H^{^{\prime }}}{H}+\frac{1}{Ha}\frac{c^{\prime }(a)}{c}\right)-\frac{v_{s}^{2}}{H^2 a^{4}}\partial^{2}\delta -\frac{%
4\pi G\bar{\mu}}{H^{2}a^{2}}\delta =0.
\end{equation}

Comparing to the same equation in Newtonian cosmology \cite{TSAG}, we find
additional term $\frac{\dot{c}}{c}$ which is due to variation of constants.

\end{document}